\documentclass[aps,preprintnumbers,amsmath,amssymb,superscriptaddress]{revtex4}

\usepackage{amsmath}
\usepackage{amssymb}
\usepackage{bm}
\usepackage{graphicx}
\usepackage{multirow}

\usepackage{textcomp}

\allowdisplaybreaks[1]

\begin{document}


\title{$D$ meson mixing as an inverse problem}

\author{Hsiang-nan Li}
\affiliation{Institute of Physics, Academia Sinica,
Taipei, Taiwan 115, Republic of China}
\author{Hiroyuki Umeeda}
\affiliation{Institute of Physics, Academia Sinica,
Taipei, Taiwan 115, Republic of China}
\author{Fanrong Xu}
\affiliation{Department of Physics, Jinan University, Guangzhou 510632, People’s Republic of China}
\author{Fu-Sheng Yu}
\affiliation{School of Nuclear Science and Technology, Lanzhou University,
Lanzhou 730000, People’s Republic of China}

\date{\today}

\begin{abstract}
We calculate the parameters $x$ and $y$ for the $D$ meson mixing in the 
Standard Model by considering a dispersion relation between them. The 
dispersion relation for a fictitious charm quark of arbitrary mass squared 
$s$ is turned into an inverse problem, via which the mixing parameters at 
low $s$ are solved with the perturbative inputs $x(s)$ and $y(s)$ from large 
$s$. It is shown that nontrivial solutions for $x$ and $y$ exist, whose values 
around the physical charm scale agree with the data in both CP-conserving and 
CP-violating cases. We then predict the observables $|q/p|-1\approx 2\times 10^{-4}$ and 
$Arg(q/p)\approx 6\times 10^{-3}$ degrees associated with the coefficient ratio 
for the $D$ meson mixing, which can be 
confronted with more precise future measurements. Our work represents the 
first successful quantitative attempt to explain the $D$ meson mixing parameters 
in the Standard Model.
\end{abstract}


%
%
%

\maketitle

%
%
%
How to understand the large $D$ meson mixing in the Standard Model has been a 
long-standing challenge. Previous evaluations based on box diagrams \cite{Cheng, BSS, Datta:1984jx}, and on heavy quark effective field theory \cite{Georgi:1992as, Ohl:1992sr} led to the mixing parameters $x$ and/or $y$ far below the current data. The updated inclusive analysis \cite{Golowich:2005pt}, including next-to-leading-order QCD corrections, still gave small $x\sim y\simeq 6\times 10^{-7}$. Some authors \cite{Bigi:2000wn, Falk:2001hx,Bobrowski:2010xg} claimed that higher dimensional operators, for which the strong Glashow-Iliopoulos-Maiani (GIM) suppression \cite{Glashow:1970gm} might be circumvented, yielded dominant contributions. This claim has not been verified quantitatively, which requires information on a large number of nonperturbative matrix elements. Another uncertainty in the heavy quark expansion originates from violation of the quark-hadron duality, which represents an error in the analytic continuation from deep Euclidean to Minkowskian domains. A simple phenomenological argument \cite{Jubb:2016mvq} indicated that  20\% duality violation could explain the width difference in the presence of the GIM cancellation.

On the other hand, the exclusive approach, where the $D$ meson mixing is extracted from hadronic processes, led to an enhancement by relevant long-distance effects \cite{Wolfenstein:1985ft, Donoghue:1985hh, Colangelo:1990hj,Buccella:1994nf,Kaeding:1995zx,Falk:2001hx, Falk:2004wg,Cheng:2010rv,Gronau:2012kq,Jiang:2017zwr}. Modern works along this direction, e.g., \cite{Cheng:2010rv,Jiang:2017zwr} showed that a half value of $y$ was accounted for roughly with contributions from two-body decays, albeit the difficulty in taking account of other multi-body channels. Thus, the quantitative understanding is still not attained in this data-driven approach, while the order of magnitude of the mixing parameters was properly described.

The complexities are attributed to the notorious difficulty of charm physics: the charm scale is too heavy to apply the chiral perturbation theory and possibly too light to apply the heavy quark expansion. Moreover, the $D$ meson mixing, strongly suppressed by the GIM mechanism, is sensitive to
nonperturbative SU(3) breaking effects \cite{Kingsley:1975fe} characterized by the strange and down quark 
mass difference, and to CKM-suppressed diagrams with bottom quarks in the loop.
On the contrary, the heavy quark expansion accommodates the data for the 
$B_{d,s}$ meson mixings satisfactorily \cite{Artuso:2015swg, Jubb:2016mvq}.


In this letter we will analyze the $D$ meson mixing in a novel approach based on a
dispersion relation, which relates $x$ and $y$ for a fictitious $D$ meson of
an arbitrary mass. The dispersion relation is separated into a low mass piece and
a high mass piece, with the former being treated as an unknown, and the latter 
being input from reliable perturbative results. We then turn the study of the $D$ 
meson mixing into an inverse problem: the mixing parameters at low mass
are solved as "source distributions", which produce the "potential"
observed at high mass. It will be demonstrated 
that nontrivial correlated solutions for $x$ and $y$ exist, 
whose values around the physical charm quark mass $m_c\approx 1.3$ GeV match 
the data in both cases with and without CP violation. Our observation 
implies that resonance properties can be extracted from asymptotic QCD by solving an inverse problem. 

Consider the analytical transition matrix element for a $D$ meson formed 
by a fictitious charm quark of invariant mass squared $s$,
\begin{eqnarray}
M_{12}(s)-\frac{i}{2}\Gamma_{12}(s)=\langle D^0(s)|{\cal H}_w^{\Delta C=2}|\bar D^0(s)\rangle,
\label{tran}
\end{eqnarray}
whose branch cut runs from the threshold $s=4m_\pi^2$ to infinity with the pion mass $m_\pi$. The effective weak Hamiltonian ${\cal H}_w^{\Delta C=2}$ contains two four-fermion operators $(V-A)(V-A)$ and $(S-P)(S-P)$, which will be abbreviated to $V-A$ and $S-P$ below, respectively.
The right hand side of Eq.~(\ref{tran}) starts with the evaluation of box diagrams, whose dispersive and absorptive contributions give rise to $M_{12}$ and $\Gamma_{12}$, respectively.
The dispersive part $M_{12}$ and the absorptive part $\Gamma_{12}$ then obey the dispersion relation \cite{Falk:2004wg}
\begin{eqnarray}
M_{12}(s)=\frac{P}{2\pi}\int_{0}^\infty ds'\frac{\Gamma_{12}(s')}{s-s'},\label{dis}
\end{eqnarray}
where $P$ denotes the principal value prescription, and the lower bound of the 
integration variable $s'$, being of $O(m_\pi^2)$, has been approximated by zero.
Equation~(\ref{tran}) governs the time evolution of the $D^0$ and $\bar D^0$ mesons, whsoe diagonalization   
yields the mass eigenstates $D_{1,2}=pD^0\pm q\bar D^0$ as linear combinations of the 
weak eigenstates $D^0$ and $\bar D^0$. The mass and width differences of $D_{1,2}$ define the mixing parameters
\begin{eqnarray}
x\equiv \frac{m_1-m_2}{\Gamma}=\frac{2M_{12}}{\Gamma},\;\;\;\;
y\equiv \frac{\Gamma_1-\Gamma_2}{2\Gamma}=\frac{\Gamma_{12}}{\Gamma},
\end{eqnarray} 
in the CP-conserving case with the total decay width $\Gamma$. 
\begin{figure}
\includegraphics[scale=0.47]{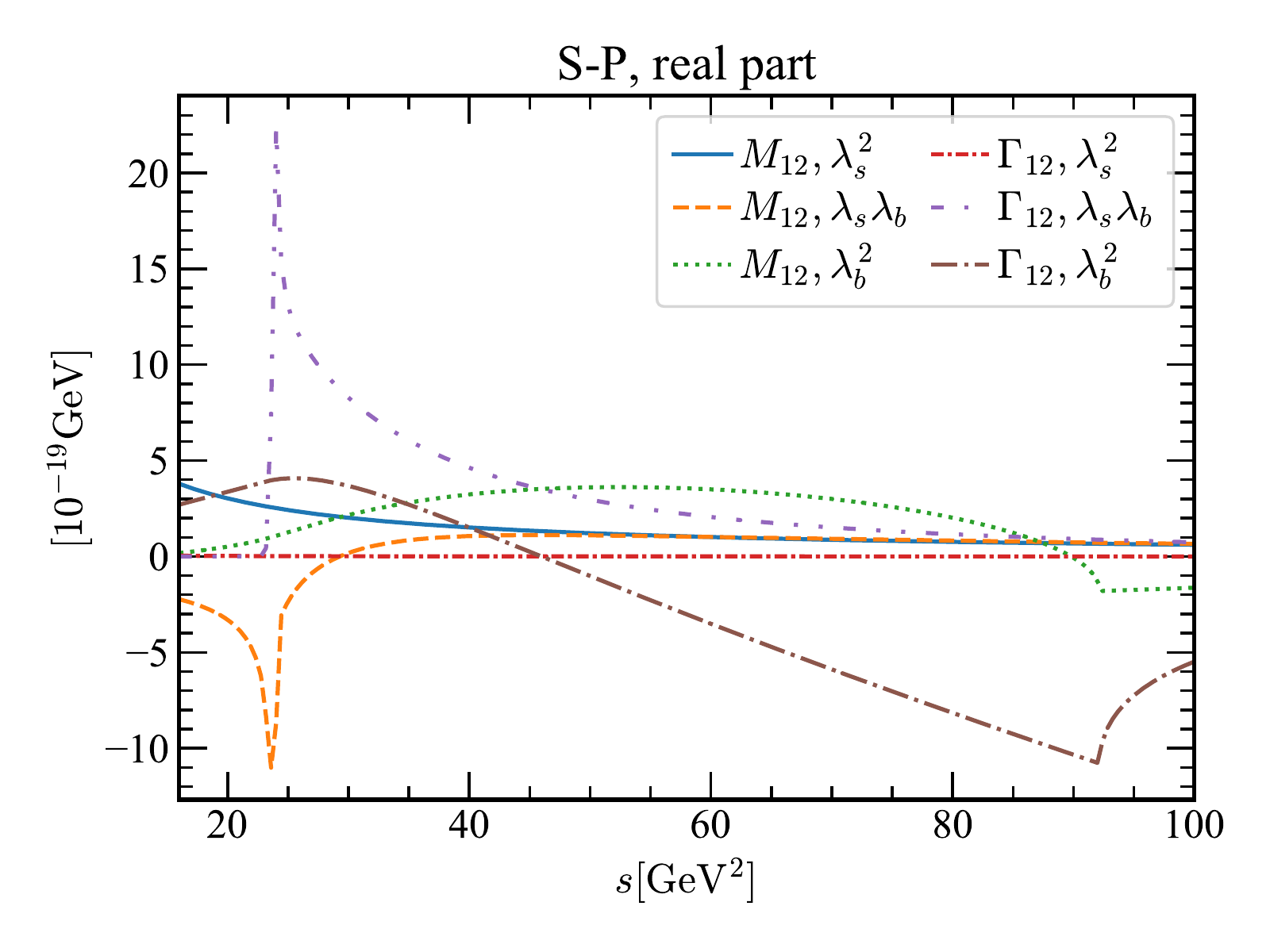}
\includegraphics[scale=0.47]{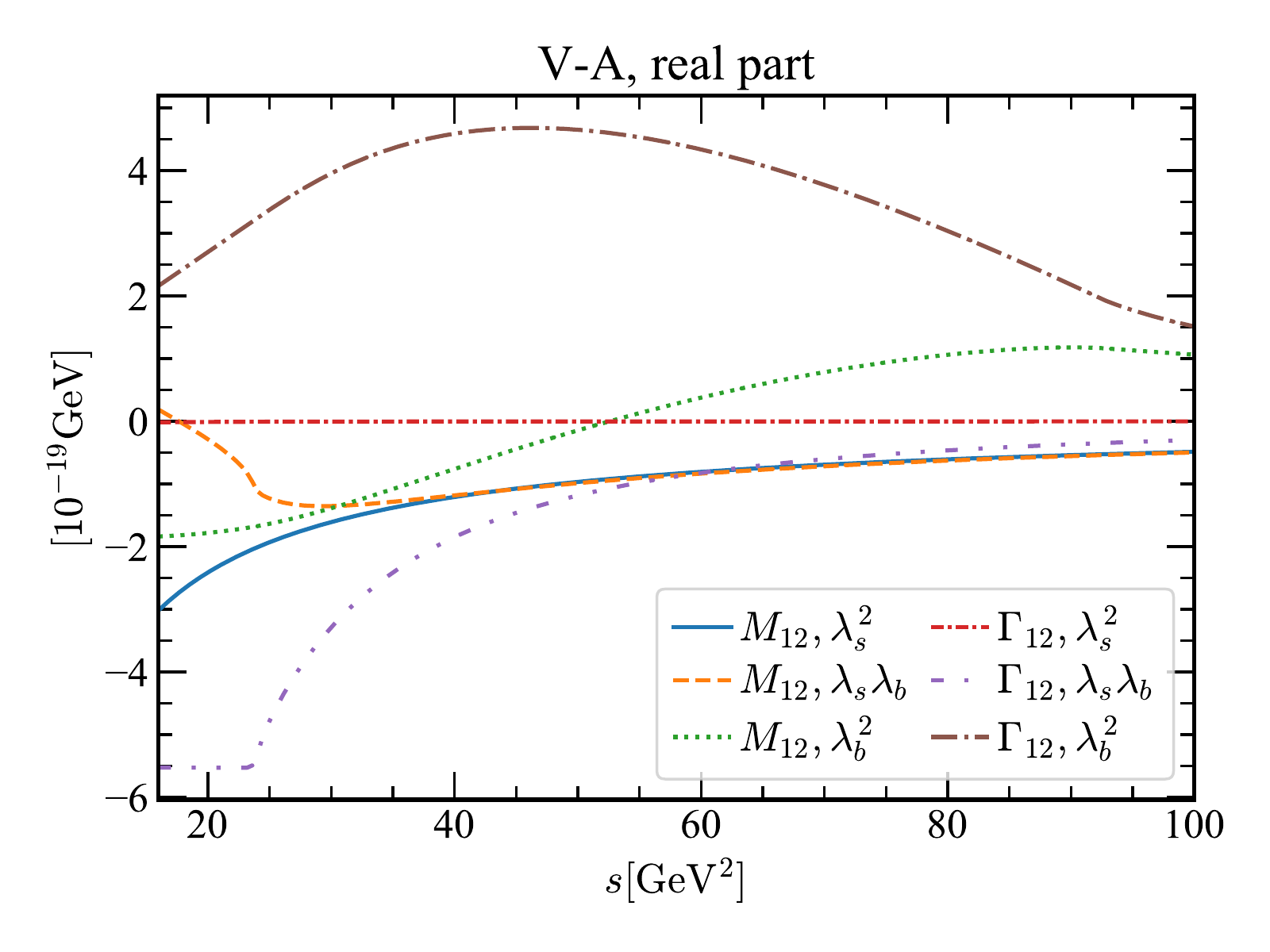}
\caption{\label{fig1}
$\lambda_s^2$, $\lambda_s\lambda_b$ and $\lambda_b^2$ contributions to the real parts
of $M_{12}$ and $\Gamma_{12}$ from the $S-P$ and $V-A$ operators.}
\end{figure}
The elements $M_{12}$ and $\Gamma_{12}$, extracted from the evaluation of box
diagrams \cite{Cheng,BSS}, can be applied to the mixing of a heavy meson with arbitrary
mass, and will be adopted directly below. The $b$ quark mass $m_b$ should remain constant 
in the evaluation of $\Gamma_{12}$, so that the fictitious $D$ meson can decay into a $b$ quark, as its 
mass crosses the $b$ quark threshold. The right hand side of Eq.~(\ref{dis}) then 
contains heavy quark contributions to be consistent with the heavy quark 
dynamics involved in $M_{12}$. The $V-A$ contribution $\Gamma_{12}^{V-A}$ is given, 
for $s>4m_b^2$, by
\begin{eqnarray}
\Gamma_{12}^{V-A}
\propto \lambda_{s}^2(B_{dd}^{(a)}-2B_{ds}^{(a)}+B_{ss}^{(a)})
+2\lambda_s\lambda_b(B_{dd}^{(a)}-B_{ds}^{(a)}-B_{db}^{(a)}+B_{sb}^{(a)})
+\lambda_{b}^2(B_{dd}^{(a)}-2B_{db}^{(a)}+B_{bb}^{(a)}),\label{CKM}
\end{eqnarray}
where $\lambda_k\equiv V_{ck}V^*_{uk}$, $k=s,b$, are the products of the
Cabibbo-Kobayashi-Maskawa (CKM) matrix elements, and the functions $B_{ij}^{(a)}$ \cite{BSS}
with the internal quarks $i,j=d,s,b$ arise from the absorptive contributions of the box diagrams 
for Eq.~(\ref{tran}). The terms up to $B_{ss}^{(a)}$ ($B_{db}^{(a)}$, $B_{sb}^{(a)}$) 
are kept in the range $s < (m_b+m_d)^2$ ($(m_b+m_d)^2\leq s<(m_b+m_s)^2$, $(m_b+m_s)^2\leq s<4m_b^2$).  
The expression of the $S-P$ contribution $\Gamma_{12}^{S-P}$ is
similar but with $B_{ij}^{(a)}$ in Eq.~(\ref{CKM}) being replaced by $C_{ij}^{(a)}$ \cite{BSS}. 
Equation~(\ref{CKM}) shows clearly that the $D$ meson mixing results from the
flavor symmetry breaking. We have confirmed that $\Gamma_{12}$ deceases like $1/s^2$ at 
large $s$, so the integral on the right hand side of Eq.~(\ref{dis}) converges.

\begin{figure}
\includegraphics[scale=0.47]{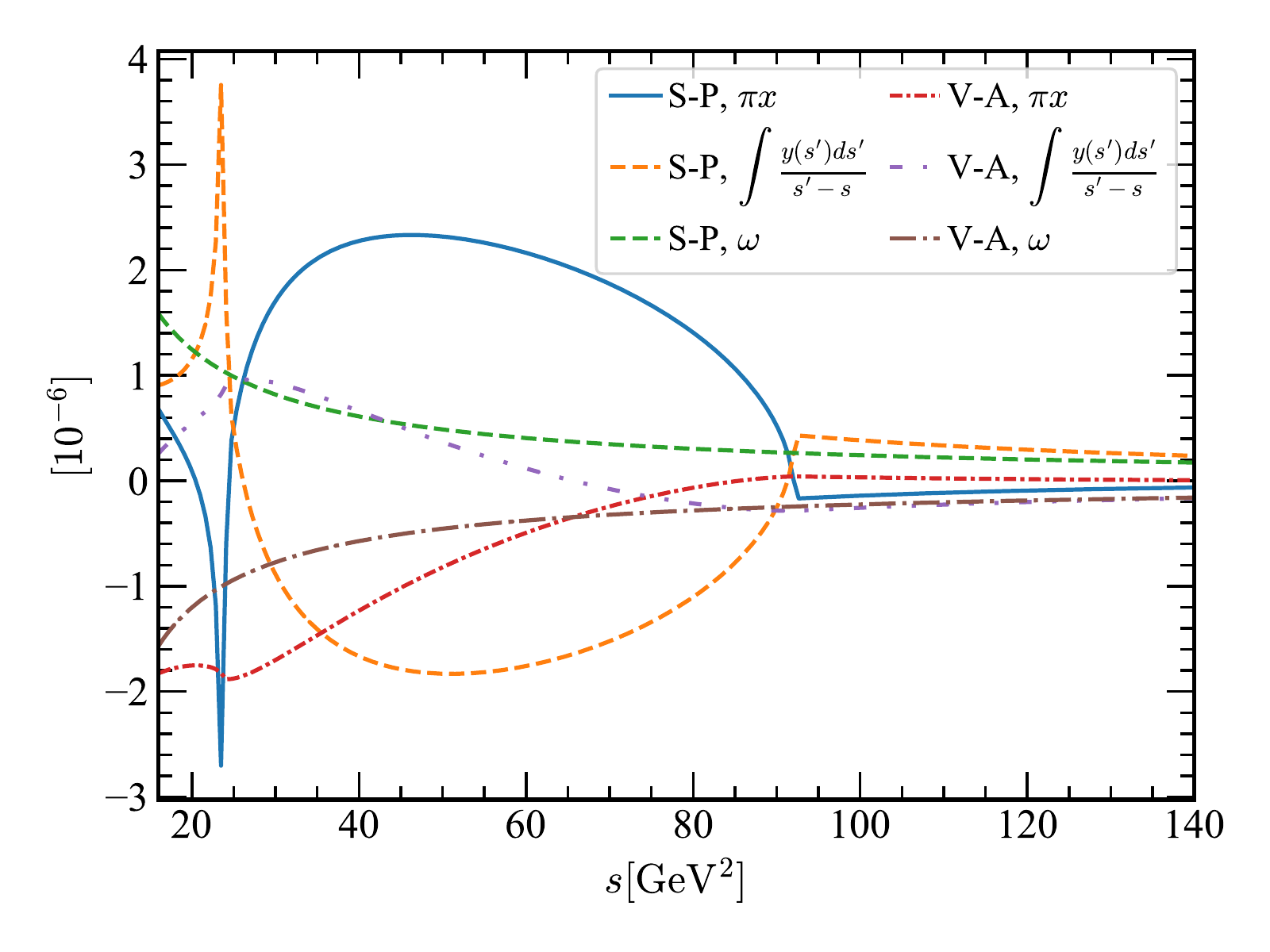}
\caption{\label{fig2}
Dispersive and absorptive contributions to the real part
of $\omega(s)$ from the $S-P$ and $V-A$ operators.}
\end{figure}

We rewrite the dispersion relation as
\begin{eqnarray}
\int_{0}^{\Lambda} ds'\frac{y(s')}{s-s'}=
\pi x(s)-\int_{\Lambda}^{\infty} ds'\frac{y(s')}{s-s'}\equiv \omega(s),\label{disxy}
\end{eqnarray}
where both sides have been divided by the measured total width 
$\Gamma_{\rm exp}=1.61 \times 10^{-12}$ GeV \cite{PDG} to
get the variables $x$ and $y$. The separation scale $\Lambda$ is arbitrary,
but should be large enough to justify the perturbative calculation of
$y$ on the right hand side, and below the $b$ quark threshold to avoid the $b$ quark
contribution to the left hand side. The product $f_D^2 m_D$ appearing in the expressions
of $M_{12}$ and $\Gamma_{12}$ \cite{BSS} on the right hand side of Eq.~(\ref{disxy}), 
with the $D$ meson decay constant $f_D$ and its mass $m_D$, 
scales like a constant in the heavy quark limit. Here we adopt the value for a $B_s$ meson \cite{PDG},
ie., $f_D^2 m_D\sim 0.3$ GeV$^3$. The behaviors of $M_{12}(s)$ and $\Gamma_{12}(s)$ 
from the $S-P$ and $V-A$ operators with the masses $m_d=5$ MeV, $m_s=109.9$ MeV, $m_b=4.8$ GeV and 
$m_W=80.379$ GeV, the separation scale $\Lambda=m_b^2/2\approx 12$ GeV$^2$, and the bag parameters 
equal to unity are displayed in Fig.~\ref{fig1}, which have been decomposed into 
three pieces proportional to the real parts of $\lambda_s^2$, $\lambda_s\lambda_b$ and $\lambda_b^2$. 
The above choice of $\Lambda$ can be regarded as being of $O(m_b^2)$, so
the perturbation theory is applicable to the mixing of the fictitious $D$ meson for 
$s>\Lambda$.  The choice of $m_s=109.9$ MeV is within the range of the strange quark mass 
$m_s=108^{+13}_{-6}$ MeV given for the 
renormalization scale $\mu=m_c$ in \cite{PDG}. It is seen in Fig.~\ref{fig2} that both 
terms on the right-hand side of Eq.~(\ref{disxy}) exhibit cusps as $s$ crosses
the $b$ quark and $b$ quark pair thresholds. Their sum $\omega(s)$ behaves smoothly
and, furthermore, turns out to be independent of $m_b$. This feature, existent for the two
four-fermion operators, indicates that $y$ in the low mass region $s<\Lambda$ 
decouples from the $b$ quark dynamics as expected.

In principle, we can have separate dispersion relations associated with the three 
CKM products. However, it is reasonable to combine all the terms in 
Eq.~(\ref{CKM}) into a single dispersion relation due to the dominance of the 
$\lambda_s^2$ contribution to the real part of $\omega(s)$. The $S-P$ and $V-A$ contributions 
are opposite in sign, and the corresponding bag parameters are roughly equal. The 
significant cancellation between these two pieces causes sensitivity to the bag parameters, 
which have not yet been computed precisely enough in lattice QCD. To reduce 
the sensitivity to this potential cancellation, we consider 
separate dispersion relations for these two operators. Equation~(\ref{disxy})
will be treated as an inverse problem, in which $\omega(s)$ for 
$s>\Lambda$ from Fig.~\ref{fig2} is an input, and $y(s)$ in the range 
$s<\Lambda$ is solved with the boundary condition $y(0)=0$ and 
the continuity of $y$ at $s=\Lambda$. That is, the "source distribution" $y(s)$ 
will be inferred from the "potential" $\omega(s)$ observed outside the distribution. 

For such an ill-posed inverse problem, the ordinary discretization method to solve an integral 
(Fredholm) equation does not work. The discretized version of Eq.~(\ref{disxy}) is 
in the form $\sum_i A_{ij}y_j=\omega_i$ with $A_{ij}\propto 1/(i-j)$. It is easy to 
find that any two adjacent rows of the matrix $A$ approach to each other as the grid 
becomes infinitely fine. Namely, $A$ tends to be singular, and has no inverse. We stress 
that this singularity, implying no unique solution, should be appreciated actually. If
$A$ is not singular, the solution to Eq.~(\ref{disxy}) will be unique, which must be
the tiny perturbative result obtained in the literature. It is the existence of multiple
solutions that allows possibility to explain the observed large $D$ meson mixing.

\begin{figure}
\includegraphics[scale=0.40]{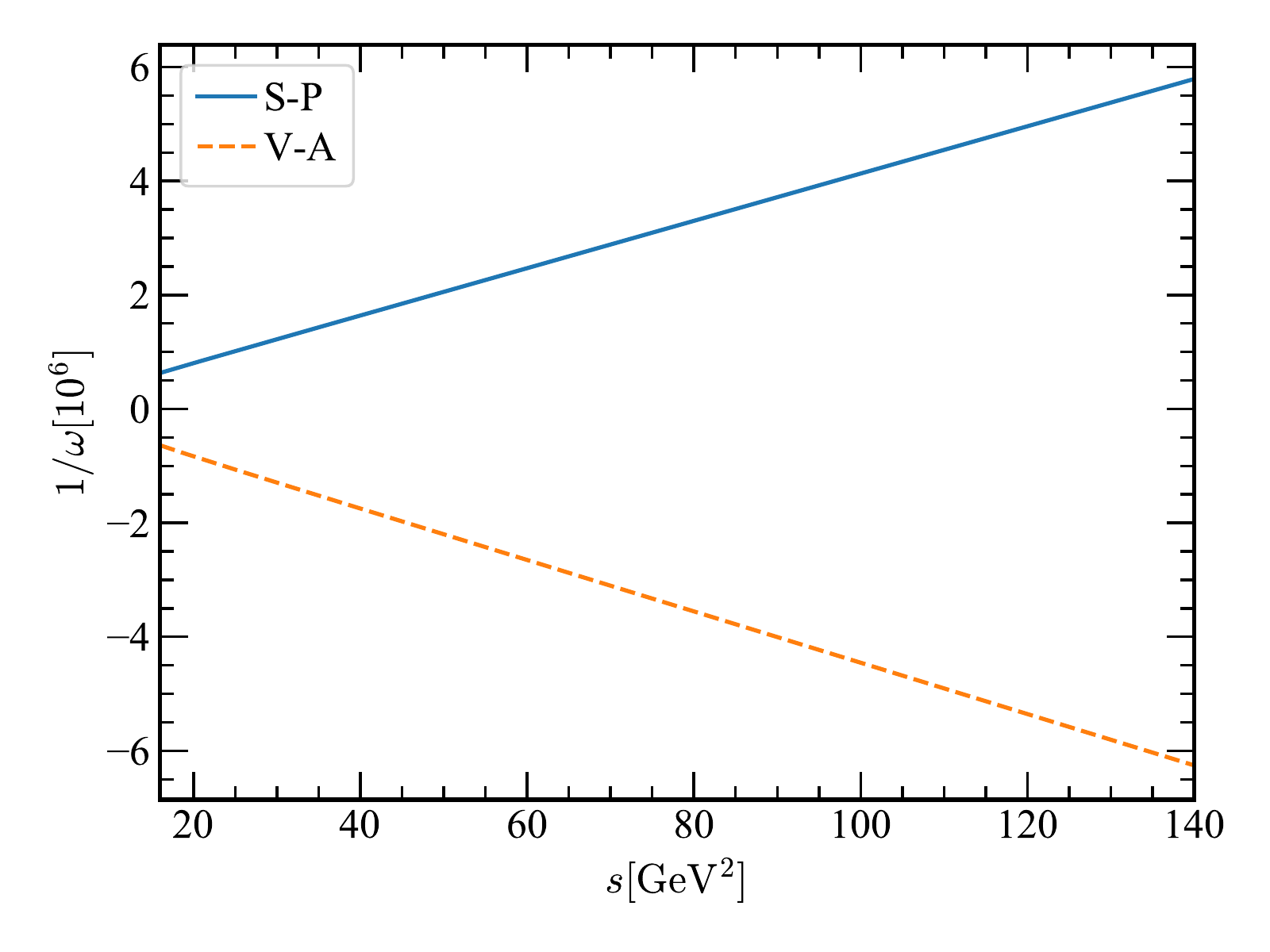}
\caption{\label{fig2b}
$s$ dependence of $1/\omega$.}
\end{figure}

We notice that the smooth curves of $\omega(s)$ can be well described by
simple functions proportional to $1/(s-m^2)$, as indicated by the
almost straight lines for $1/\omega(s)$ down to $s=15$ GeV$^2$ in 
Fig.~\ref{fig2b}. These straight lines, as extrapolating to the low $s$ region, 
cross the horizontal axis at some small scale $s=m^2$. 
The power-law behavior is understandable, 
since only the effect from the monopole component of the distribution 
dominates at large $s$, which decreases like $1/s$. 
The meaning of the scale $m^2$ will become clear later.
If $\omega(s)$ followed the power law exactly, the solution to
Eq.~(\ref{disxy}) would be a $\delta$-function, $y(s)\propto \delta(s-m^2)$.
The slight deviation from the power-law behavior suggests mild broadening of $y(s)$ into 
a resonance-like distribution located at $s\approx m^2$, if $m^2>0$.

Viewing the difficulty to solve an inverse problem with
multiple solutions and the qualitative resonance-like behavior of a solution, 
we propose the parametrization
\begin{eqnarray}
y(s)=\frac{Ns[b_0+b_1(s-m^2)+b_2(s-m^2)^2]}{[(s-m^2)^2+d^2]^2},\label{para}
\end{eqnarray}
for $0\le s\le \Lambda$, and determine the free parameters $b_0$, $b_1$, $b_2$, 
$m^2$ and $d$ from the best fit to the input $\omega(s)$. The normalization $N$ 
respects $Ns/[(s-m^2)^2+d^2]^2\to \delta(s-m^2)$ in the vanishing width
limit $d\to 0$. Equation~(\ref{para}) with the completely free
parameters is general enough, which can also describe a nonresonant 
behavior with $m^2 < 0$ and a flat behavior with large $d$.  
The convergence of the expansion in the numerator will be verified, so 
keeping terms up to $(s-m^2)^2$ is sufficient. Equation~(\ref{para})
obeys the boundary condition $y(0)=0$. The continuity of $y(s)$ at $s=\Lambda$,
ie., the equality of $y(\Lambda)$ to the perturbative input
imposes a constraint among the five parameters. We emphasize that a systematic
expansion of $y(s)$ in terms of a
complete basis of orthogonal functions also works, but the numerical analysis is
more tedious, and will be performed in a forthcoming paper. 

\begin{figure}
\includegraphics[scale=0.4]{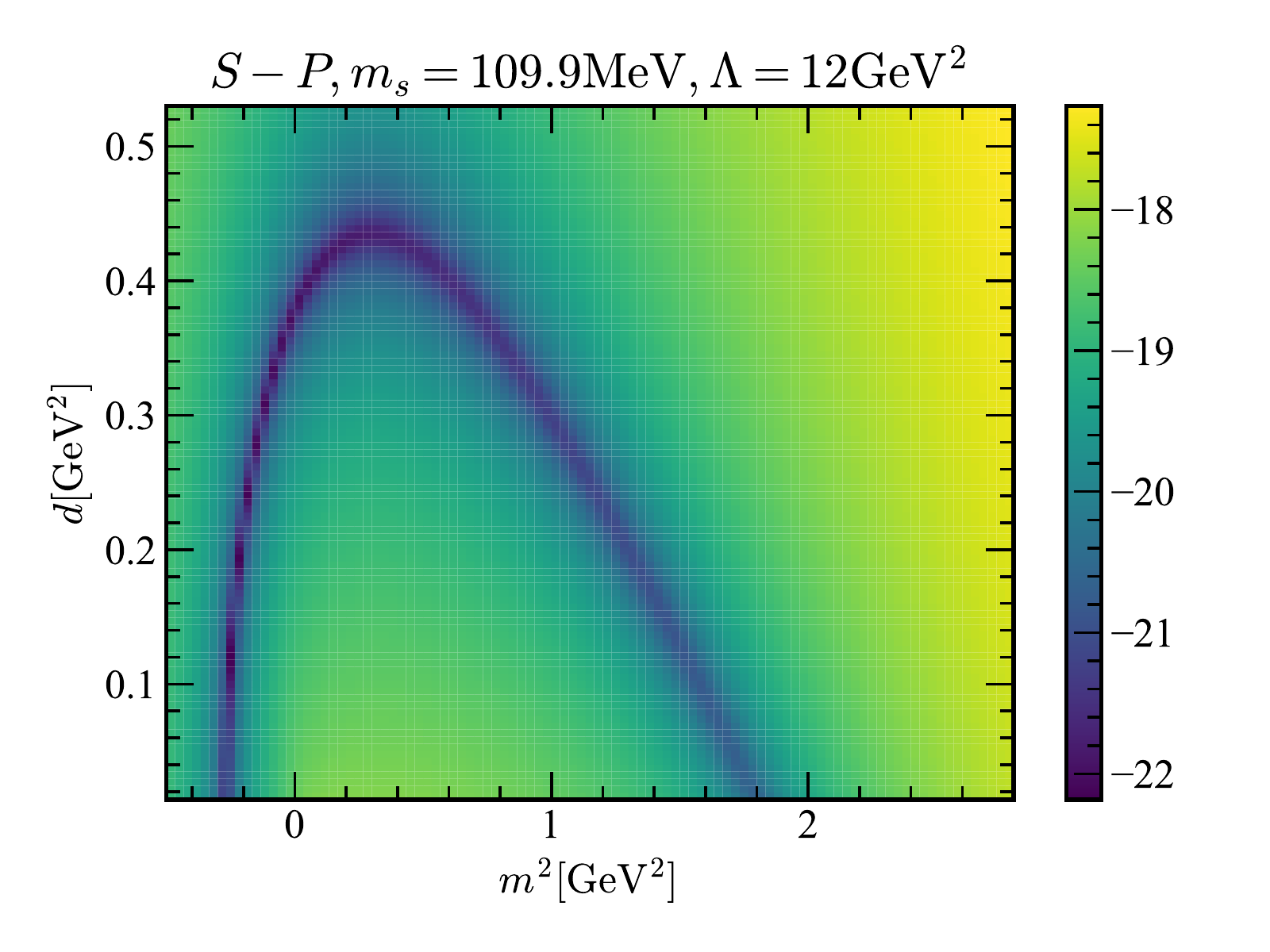}
\includegraphics[scale=0.4]{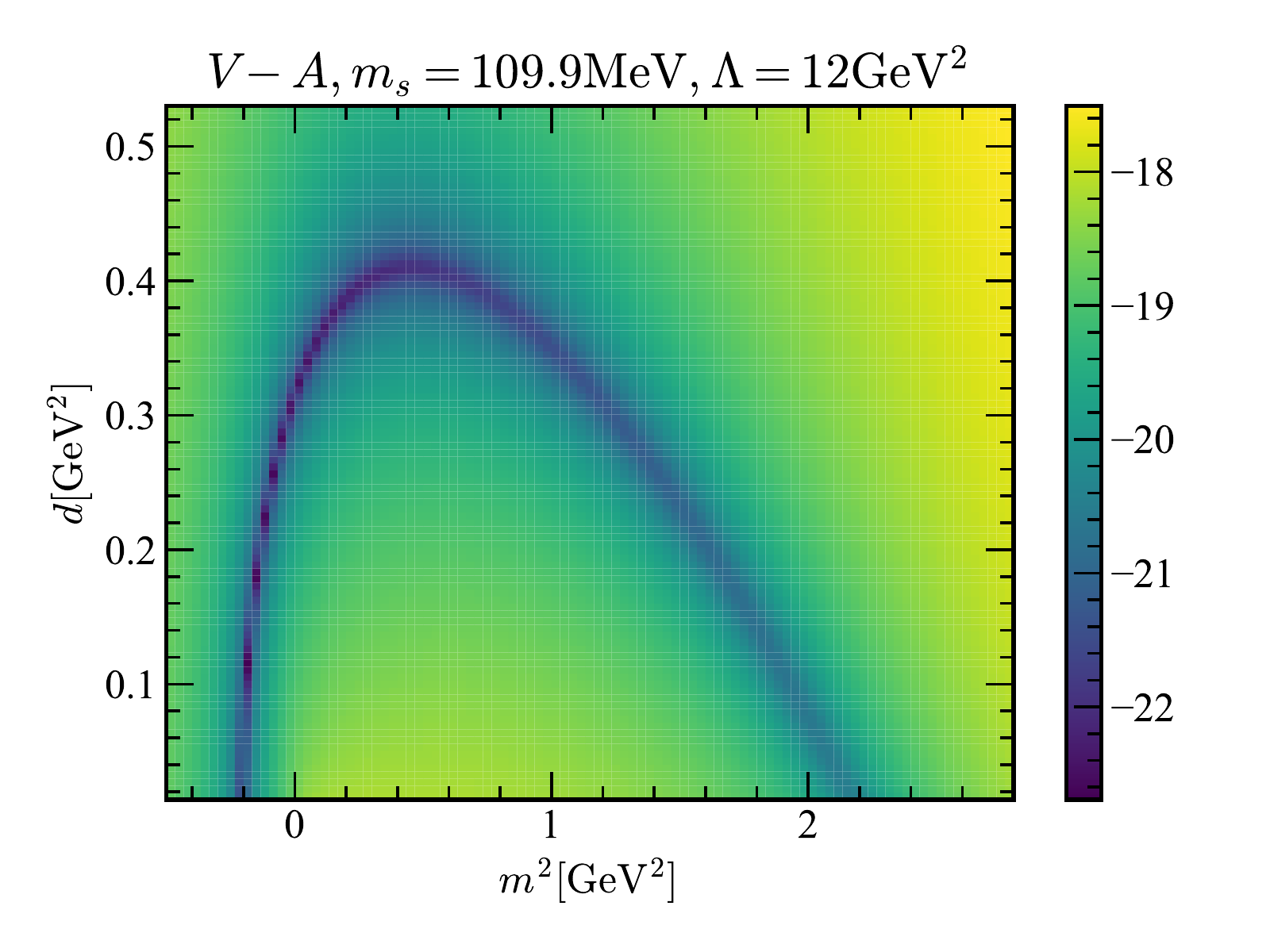}
\caption{\label{fig3}
Distributions of GOF minima in the common logarithmic scale on the $m^2$-$d$ plane.}
\end{figure}

The separation scale $\Lambda$ introduces an end-point
singularity to the integral on the right hand side of Eq.~(\ref{disxy}),
as $s\to\Lambda$. To reduce the effect caused by this artificial singularity,
we consider $\omega(s)$ from the range 30 GeV$^2<s<$ 250 GeV$^2$, in which
200 points $s_i$ are selected. 
We have checked the cases with 100, 200 and 300 points, and confirmed that the results have 
little dependence on these numbers.
For each point $(m^2,d)$ on the $m^2$-$d$ plane, we search for $b_0$ and $b_1$,
that minimize the deviation
\begin{eqnarray}
\sum_{i=1}^{200} \left|\int_{0}^{\Lambda} ds'\frac{y(s')}{s_i-s'}-\omega(s_i)\right|^2.\label{dev}
\end{eqnarray}
The above definition, characterizing the relative quality of solutions, is referred to as the 
goodness-of-fit (GOF) hereafter. The value of $b_2$ is fixed by the continuity constraint at 
$s=\Lambda$. The scanning on the $m^2$-$d$ plane generates the arc-shaped 
distribution of the GOF minima associated with the $S-P$ operator in Fig.~\ref{fig3}, which ranges 
roughly in -0.2 GeV$^2<m^2<1.8$ GeV$^2$. The minima along the arc, 
having similar GOF about $10^{-21}$-$10^{-22}$ relative to 
$10^{-17}$ from outside the arc, hint the existence of multiple solutions.
If a resonance-like solution with $m^2 \sim m_c^2$ and small $d$ exists, ie., obeys 
the dispersion relation, it will be revealed by the scanning, and indeed it is as shown
in Fig.~\ref{fig3}.

\begin{figure}
\includegraphics[scale=0.4]{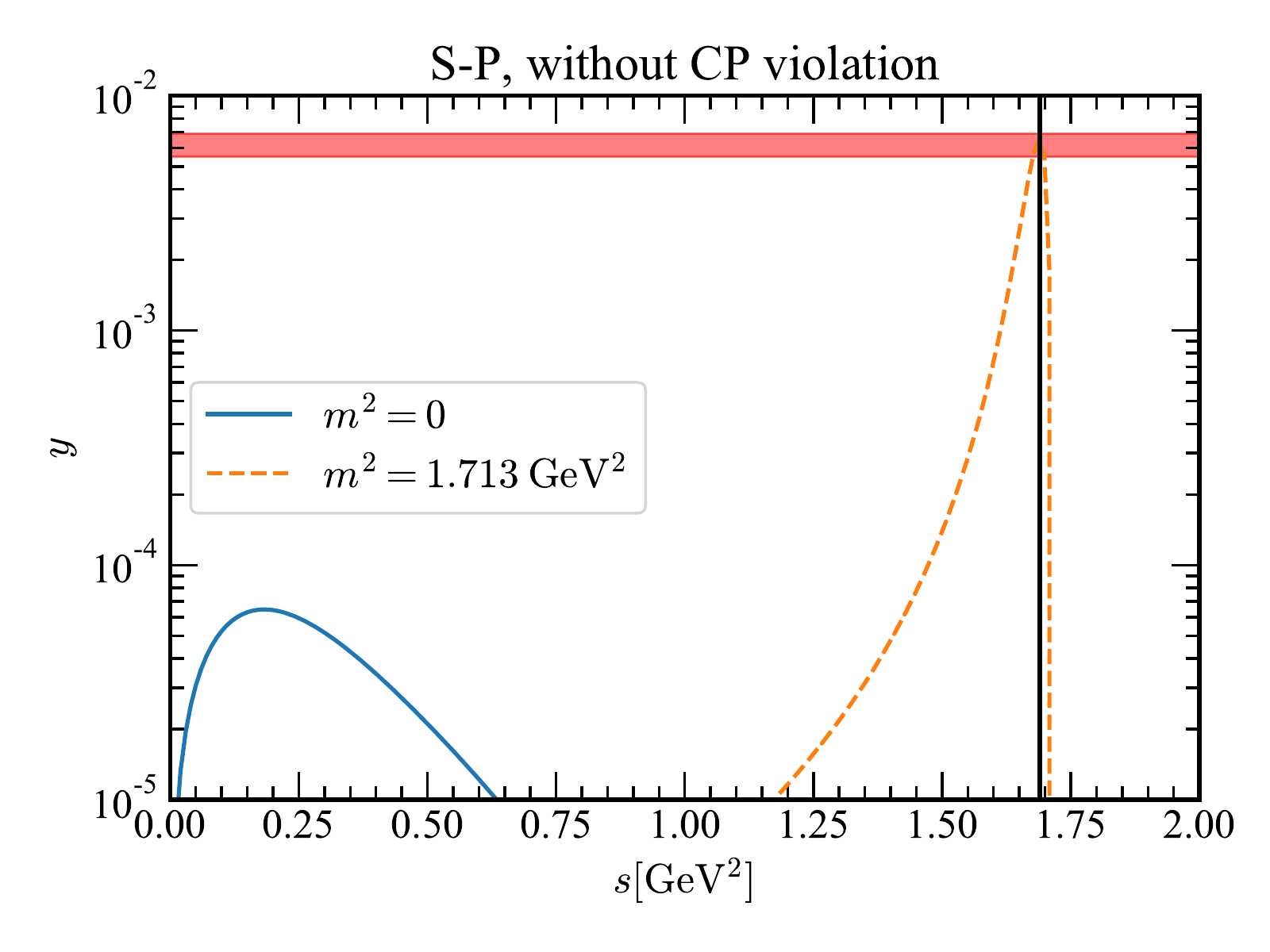}
\caption{\label{figure}
Comparison of a typical nonresonant solution and a resonance-like solution for $y$. The horizontal band
represents the data with $1\sigma$ errors, and the vertical line corresponds to $s=m_c^2$.}
\end{figure}

Evaluating $y(s)$ at low $s$ in perturbation theory with a finite running coupling 
constant $\alpha_s$, we get different results at various orders. These different results 
lead to almost identical $\omega(s)$ in the large $s$ limit, where $\alpha_s$ diminishes. 
The solutions from $m^2$ away from $m_c^2$ might correspond to fixed-order results, since 
they generate tiny $y$ at the physical scale $m_c^2$, while those near $m^2\approx m_c^2$ 
correspond to nonperturbative results. We select a typical nonresonant solution for $y(s)$ 
with $m^2=0$ and $d=0.38$ GeV$^2$ from the arc associated with the $S-P$ operator, and 
compare it to a resonance-like solution with $m^2= 1.713$ GeV$^2$ and 
$d= 3.876\times 10^{-2}$ GeV$^2$ from the same arc in Fig.~\ref{figure}. The dramatic distinction
in the shape and in the order of magnitude between these two cases supports
that Eq.~(\ref{para}) is general enough to exhibit very different behaviors. The observation
that the above perturbative and nonperturbative solutions give the same $\omega(s)$ at 
large $s$ realizes the concept of the global
quark-hadron duality postulated in QCD sum rules \cite{Shifman:2000jv}. 
The arc-shaped distribution from the $V-A$ operator is also displayed in Fig.~\ref{fig3},
where a solution with $m^2\approx m_c^2$ has a large $d$, so its contribution to
$y$ is negligible.
 
\begin{figure}
\includegraphics[scale=0.4]{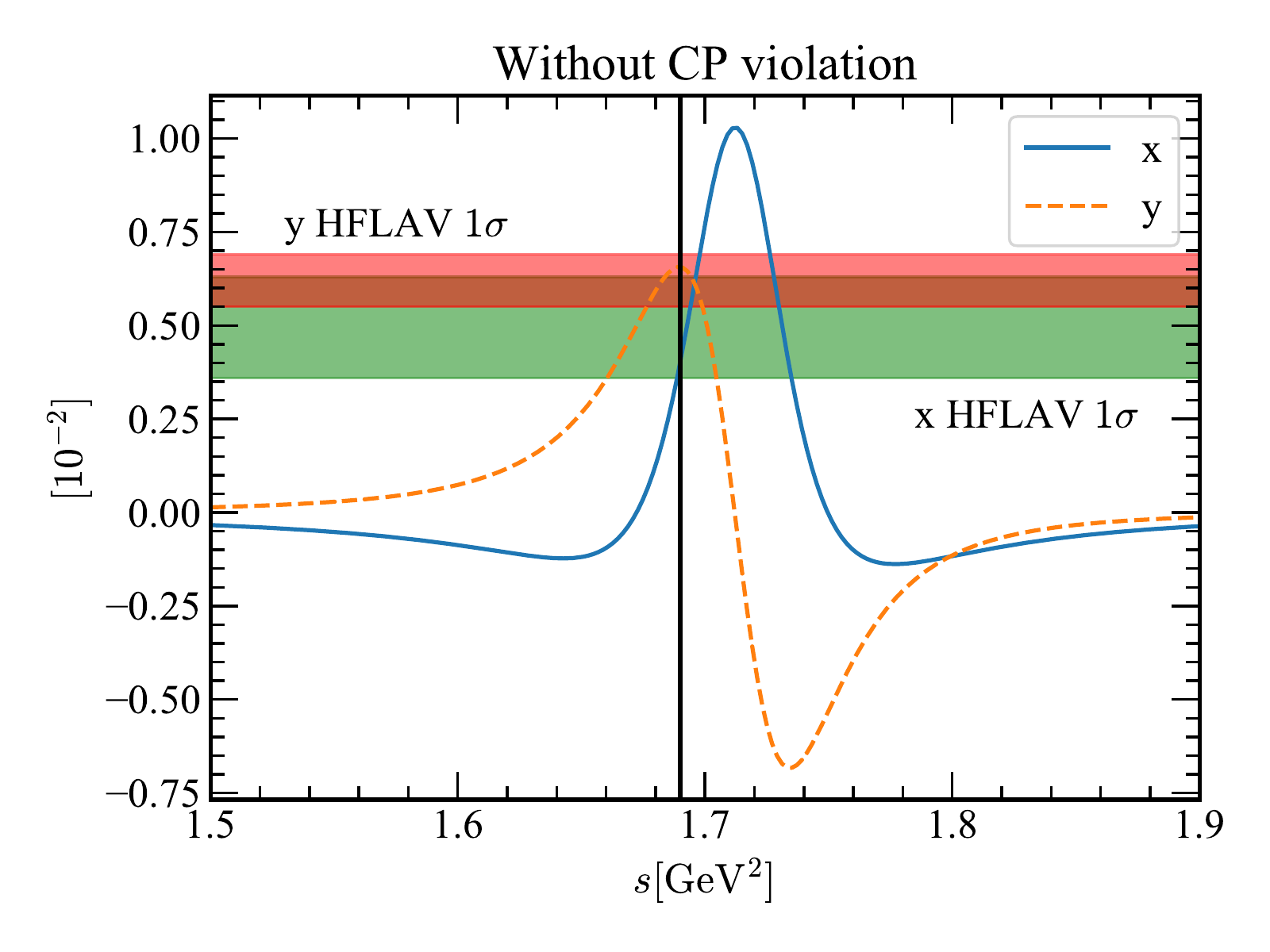}
\includegraphics[scale=0.4]{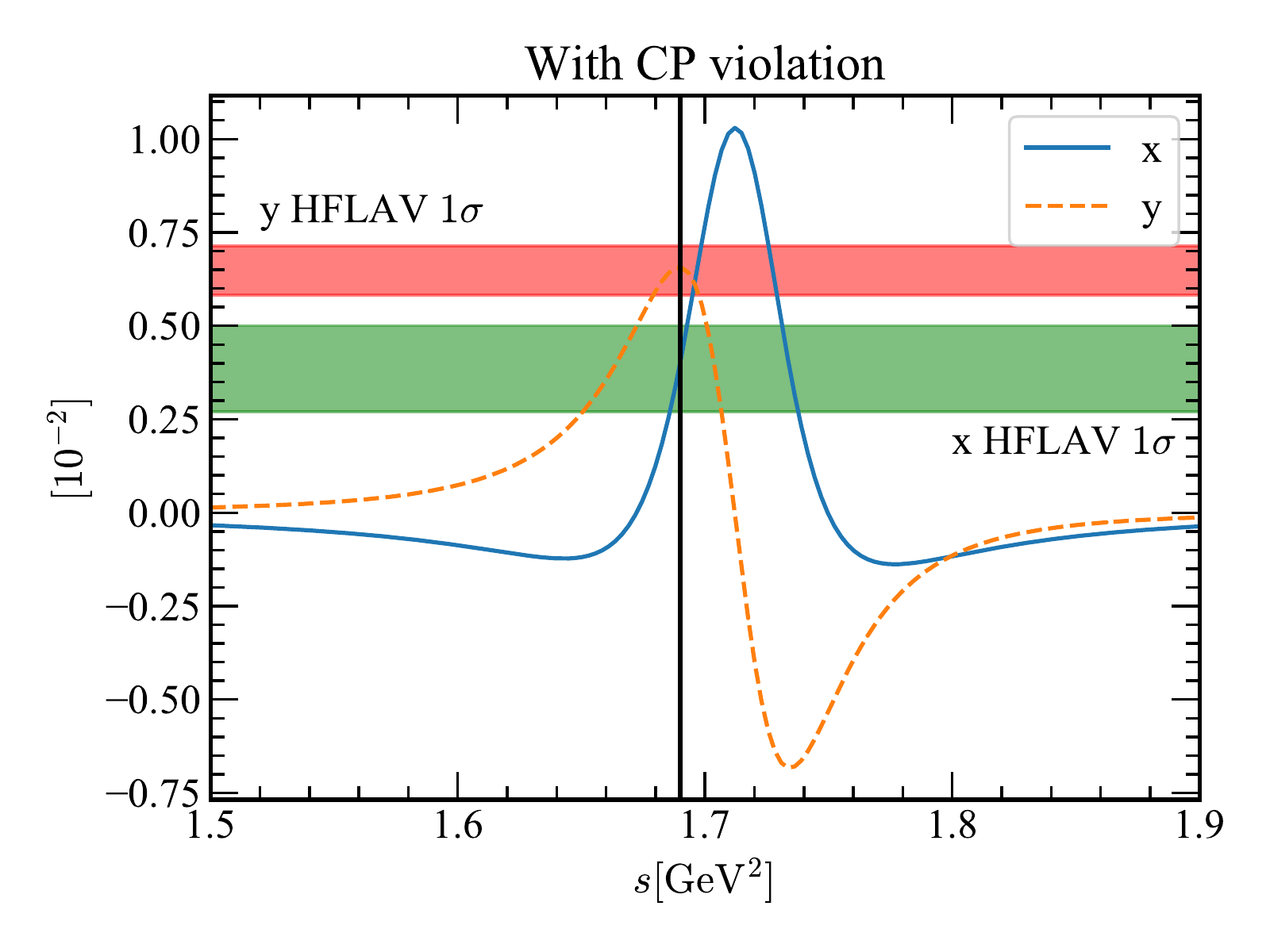}
\caption{\label{fig4}
$x(s)$ and $y(s)$ in the cases without and with CP violation.}
\end{figure}

Selecting a point $(m^2,d)$ on the arc, we get a solution of $y(s)$. Substituting 
the obtained $y(s)$, ie., $\Gamma_{12}(s)$ in the whole range of $s$ into  
Eq.~(\ref{dis}), we calculate the corresponding $x(s)$. The values $x(m_c^2)$
and $y(m_c^2)$ are then compared with the data. 
It is seen in the left plot of Fig.~\ref{fig4} that the data $x=(0.50^{+0.13}_{-0.14})\%$ 
and $y=(0.62\pm 0.07)\%$ in the CP-conserving case \cite{Amhis:2019xyh} 
can be accommodated simultaneously by the $S-P$ contribution with the parameters 
\begin{eqnarray}
& &m^2=  1.713\;{\rm GeV}^2,\;\;d= 3.876\times 10^{-2}\;{\rm GeV}^2,\nonumber\\
& &b_0= -3.296\times 10^{-5}\;{\rm GeV}^2,\;\;
b_1=-3.234\times 10^{-2},\;\;
b_2=5.617\times 10^{-2}\;{\rm GeV}^{-2}.\label{bfit}
\end{eqnarray}
Equation~(\ref{bfit}) justifies that the lower bound of the integral in Eq.~(\ref{dis}), being
of $O(m_\pi^2)$, can be set to zero safely, because $y(s)$ takes substantial values only 
around $s\sim O(m_c^2)$. We remind that the values in Eq.~(\ref{bfit})
are representative, and their slight variations are allowed for explaining the data of $x$ and $y$
within 1$\sigma$. For instance, the width $d$ is allowed to vary by 20\%. 
The uncertainties from the fitting procedure and from the parametrization for $y(s)$ will be investigated 
rigorously in a subsequent publication.

It has been concluded \cite{Jiang:2017zwr} that two-body modes in $D$ meson decays
are insufficient for understanding $y$, and multi-particle modes play a crucial role for this purpose. 
When $s$ increases, single strange quark channels with destructive contributions, like $KKK\pi$, are 
enhanced by phase space, and double strange quark channels with constructive
contributions, like $KKKK$, are opened. This tendency fits the behavior of $y(s)$ 
in Fig.~\ref{fig4}, which first decreases from a positive value expected in the 
two-body analysis \cite{Jiang:2017zwr} to a negative value, and
then increases with $s$. It also explains why the width $d$, within which the above
oscillation occurs, is of $O(m_s^2)$. As a single resonance around $s\approx m_c^2$ accommodates
the data, it would hint that the multi-particle channels with the total rest mass around $m_c$
give dominant contributions to $y$. Certainly, our observation does not exclude 
other shorter peaks at lower $s$ but above the threshold for two-body channels. That is, 
the curve in Fig.~\ref{fig4} has caught the major 
features of $y(s)$, though its true behavior might be more complicated. 
We have also examined that the $b_1$ term dominates,
and the $b_2$ term contributes only about 10\% of $x(m_c^2)$ and $y(m_c^2)$. 
The convergence of the parametrization in Eq.~(\ref{para}) is verified.
To test whether $x(m_c^2)$ and $y(m_c^2)$ exhibit a quadratic rise with $m_s$,
as expected from the SU(3) symmetry breaking \cite{Falk:2001hx}, we fix $\Lambda =12~\mathrm{GeV}^2$ and 
$m^2= 1.713\;{\rm GeV}^2$ in Eq.~(\ref{bfit}), and then derive $x(m_c^2)$ and $y(m_c^2)$ 
from the dispersion relation for various $m_s$. The quadratic increase with a vanishing slope
at small $m_s$ is indeed observed.

As CP violation is allowed, both $M_{12}$ and $\Gamma_{12}$ become complex due to 
the weak phase in the CKM matrix elements, but Eq.~(\ref{dis}) still holds. 
The expressions of $x$ and $y$ in terms of the complex $M_{12}$ and 
$\Gamma_{12}$ are referred to \cite{PDG}. In this case the same parameters
$\Lambda=12$ GeV$^2$ and $m_s=109.9$ MeV are chosen, and the $\lambda_b^2$ contribution 
is found to dominate the imaginary part of $\omega(s)$. An additional parametrization 
similar to Eq.~(\ref{para}) but with primed parameters is proposed. The imaginary part 
of $\omega(s)$ is fitted by the primed parametrization independently of the fitting to its real part. 
The scanning on the $m^2$-$d$ planes yield the arc-shaped distributions of the GOF minima 
similar to Fig.~\ref{fig3}. Taking a common value for $m^2$ and $m^{\prime 2}$, one finds that the real 
part of the $S-P$ contribution still dominates $x$ and $y$. The parameters in Eq.~(\ref{bfit}) and 
\begin{eqnarray}
& &m^{\prime 2}=m^2=  1.713\;{\rm GeV}^2,\;\; d^\prime= 4.970\;{\rm GeV}^2,\nonumber\\
& &b^\prime_0= -8.238\times 10^{-7}\;{\rm GeV}^2,\;\;
b^\prime_1=4.355\times 10^{-7},\;\;
b^\prime_2=-7.192\times 10^{-8}\;{\rm GeV}^{-2},\label{bfittest}
\end{eqnarray}
for the $S-P$ imaginary contribution, and those for the $V-A$ contribution,
which are not presented for simplicity, accommodate the data $x=(0.39^{+0.11}_{-0.12})\%$ 
and $y=(0.651^{+0.063}_{-0.069})\%$ \cite{Amhis:2019xyh} simultaneously, as illustrated in the 
right plot of Fig.~\ref{fig4}.
Given the parameters in Eqs.~(\ref{bfit}) and (\ref{bfittest}) and those of the $V-A$ contribution,
we then derive $|q/p|-1\approx 2.2\times 10^{-4}$ and $Arg(q/p)\approx (6.2\times 10^{-3})^\circ$ 
associated with the coefficient ratio as predictions, which are comparable to the data 
$q/p=(0.969^{+0.050}_{-0.045})e^{i(-3.9^{+4.5}_{-4.6})^\circ}$ 
\cite{Amhis:2019xyh}, and can be confronted with more precise measurements in future. 

To examine the uncertainty from the theoretical input, we increase $m_s$ to, say, 130.6 MeV, 
for which $\Lambda$ needs to increase to 14 GeV$^2$ accordingly to accommodate the observed $x$ and $y$.
That is, a positive correlation between $m_s$ and $\Lambda$ is observed.
In this case, the representative parameters for the real and imaginary parts of the $S-P$ contribution 
from the fit are
\begin{eqnarray}
& &m^2 =m^{\prime 2}= 1.720~\mathrm{GeV},\;\; d = 5.467\times 10^{-2}~\mathrm{GeV}^2,\;\; 
d^\prime= 5.267~\mathrm{GeV}^2, \nonumber\\
& &b_0 = 7.989\times 10^{-6}~{\rm GeV}^2, \;\;b_1 = -3.025\times 10^{-2}, \;\;
b_2 = 3.011\times 10^{-2}~{\rm GeV}^{-2},\nonumber\\
& & b_0^\prime= -1.438\times 10^{-6}~{\rm GeV}^{2}, \;\;b_1^\prime= 6.680\times 10^{-7}, \;\;
b_2^\prime= -8.870\times 10^{-8}~{\rm GeV}^{-2}.
\end{eqnarray}
Compared to Eqs.~(\ref{bfit}) and (\ref{bfittest}), the result of $m^2$ varies slightly,
the dominant coefficient $b_1$ changes by 10\% roughly, and $d$ exhibits about 30\% uncertainty. 
The above parameters, together with those for the $V-A$ contribution, lead to $|q/p|-1\approx 3.2\times 10^{-4}$ 
and $Arg(q/p)\approx (7.1\times 10^{-3})^\circ$.
It is seen that our predictions for $|q/p|-1$ and $Arg(q/p)$ are quite stable with respect to the 
variation of $m_s$, which change by only $\sim$10\%-30\%.  That is, $q/p$ can serve as an 
ideal observable for constraining new physics effects.

This work represents the first successful quantitative attempt
in the sense that definite values have been presented for the $D$ meson mixing parameters 
$x$ and $y$ in both the CP-conserving and CP-violating cases in the Standard Model. The key is to transform the dispersion relation between $x$ and $y$ into an inverse problem, in which the nonperturbative observables at low mass are solved 
with the perturbative inputs from high mass. It is nontrivial to find a solution
under the analyticity constraint from the perturbative inputs that explains the
data of $x$ and $y$. The accommodation of the data by a single resonance around 
the charm mass hints that multi-particle channels of $D$ meson decays 
give dominant contributions to $y$. If such a solution does not exist, it would be a strong
indication that the large mixing parameters are attributed to new physics. The obtained 
solution has been employed to predict the coefficient ratio $q/p$ in the CP-violating case.
To improve the precision of our results, high-power corrections to the inputs can be included systematically. 
Theoretical uncertainties in this approach will be investigated in detail in the future. 
Once the $D$ meson mixing is understood, relevant data, especially those for the 
coefficient ratio $q/p$, can be used to constrain new physics effects appearing in the 
box diagrams. Our approach will be developed into a fundamental nonperturbative 
QCD formalism, with the insight that resonance properties are extractable from 
asymptotic QCD as demonstrated in the $D$ meson mixing case.

{\bf Acknowledgement}

This work was supported in part by MOST of R.O.C. under Grant No.
MOST-107-2119-M-001-035-MY3, and by NSFC under Grant Nos. U1932104, 11605076, 
U173210, and 11975112.


\end{document}